\documentclass{aa}

\usepackage{enumerate}

\makeatletter
\g@addto@macro{\UrlBreaks}{\UrlOrds}
\makeatother
\usepackage{CJKutf8}

\newcommand{\hg}{H$\gamma$}
\newcommand{\hd}{H$\delta$}

\newcommand{\hda}{\hd$_\mathrm{A}$}

\usepackage{import}
\usepackage{amsmath}
\usepackage{amssymb}
\usepackage{natbib}
\bibpunct{(}{)}{;}{a}{}{,}
\usepackage{lscape}
\usepackage{xcolor}
\usepackage[normalem]{ulem}
\usepackage{blindtext}
\usepackage{graphicx}
\usepackage[para]{threeparttable}
\usepackage{txfonts}
\usepackage[colorlinks=true,linkcolor=blue,citecolor=blue,urlcolor=blue]{hyperref}
\begin{document}

   \title{Less is less: photometry alone cannot predict the observed spectral indices of $z\sim1$ galaxies from the LEGA-C spectroscopic survey}

    \titlerunning{Photometry alone cannot predict the observed spectral indices of $z\sim1$ galaxies}

   \author{Angelos Nersesian 
          \inst{1}
          \and
          Arjen van der Wel
          \inst{1,2}
          \and
          Anna Gallazzi
          \inst{3}
          \and
          Joel Leja
          \inst{4,5,6}
          \and
          Rachel Bezanson
          \inst{7}
          \and
          Eric F. Bell
          \inst{8}
          \and 
          Francesco D'Eugenio
          \inst{9, 10}
          \and
          Anna de Graaff
          \inst{2}
          \and 
          Yasha Kaushal
          \inst{7}
          \and
          Marco Martorano
          \inst{1}
          \and
          Michael Maseda
          \inst{11, 12}
          \and
          Stefano Zibetti
          \inst{3}
          }

   \institute{Sterrenkundig Observatorium Universiteit Gent, Krijgslaan 281 S9, B-9000 Gent, Belgium\\
            \email{\textcolor{blue}{angelos.nersesian@ugent.be}}
        \and
            Max-Planck Institut f\"{u}r Astronomie K\"{o}nigstuhl, D-69117, Heidelberg, Germany
        \and  
            Osservatorio Astrofisico di Arcetri, Largo Enrico Fermi 5, I-50125 Firenze, Italy
        \and 
            Department of Astronomy and Astrophysics, 525 Davey Lab, The Pennsylvania State University, University Park, PA 16802, USA
        \and
            Institute for Gravitation and the Cosmos, The Pennsylvania State University, University Park, PA 16802, USA
        \and
            Institute for Computational and Data Sciences, The Pennsylvania State University, University Park, PA 16802, USA
        \and
            Department of Physics and Astronomy and PITT PACC, University of Pittsburgh, Pittsburgh, PA 15260, USA
        \and
            Department of Astronomy, University of Michigan, 1085 South University Avenue, Ann Arbor, MI 48109, USA
        \and
            Kavli Institute for Cosmology, University of Cambridge, Madingley Road, Cambridge, CB3 0HA, UK
        \and 
            Cavendish Laboratory - Astrophysics Group, University of Cambridge, 19 JJ Thomson Avenue, Cambridge, CB3 0HE, UK
        \and
            Department of Astronomy, University of Wisconsin, 475 N. Charter Street, Madison, WI 53706, USA
        \and 
            Leiden Observatory, Leiden University, P.O. Box 9513, 2300 RA, Leiden, The Netherlands
             }

   \date{Received 28 April 2023; accepted 26 October 2023}

 
  \abstract
   {}
   {We test whether we can predict optical spectra from deep-field photometry of distant galaxies. Our goal is to perform a comparison in data space, highlighting the differences between predicted and observed spectra.}
   {The Large Early Galaxy Astrophysics Census (LEGA-C) provides high-quality optical spectra of thousands of galaxies at redshift $0.6<z<1$. Broad-band photometry of the same galaxies, drawn from the recent COSMOS2020 catalog, is used to predict the optical spectra with the spectral energy distribution (SED) fitting code {\tt Prospector} and the MILES stellar library. The observed and predicted spectra are compared in terms of two age and metallicity-sensitive absorption features (\hda~and Fe4383).}
   {The global bimodality of star-forming and quiescent galaxies in photometric space is recovered with the model spectra. But the presence of a systematic offset in the Fe4383 line strength and the weak correlation between the observed and modeled line strength imply that accurate age or metallicity determinations cannot be inferred from photometry alone.}
   {For now we caution that photometry-based estimates of stellar population properties are determined mostly by the modeling approach and not the physical properties of galaxies, even when using the highest-quality photometric datasets and state-of-the-art fitting techniques. When exploring a new physical parameter space (i.e. redshift or galaxy mass) high-quality spectroscopy is always needed to inform the analysis of photometry.}

   \keywords{galaxies: high-redshift -- galaxies: photometry}

   \maketitle
%

\section{Introduction}

The spectral energy distribution (SED) of galaxies encodes a plethora of information about their unresolved stellar populations, dust properties, and physical state of gas \citep[see review by][]{Conroy_2013ARA&A..51..393C}. The observed optical component of the SED can be compared with complex stellar population synthesis (SPS) models \citep[e.g.][]{Bruzual_2003MNRAS.344.1000B, Maraston_2005MNRAS.362..799M, Conroy_2009ApJ...699..486C, Eldridge_2017PASA...34...58E, Byrne_2022MNRAS.512.5329B}, which incorporate the latest developments on stellar evolution theory. By matching observations with theory, it becomes possible to derive relevant physical quantities including the metallicity and ages of the stellar populations. Having an accurate estimate of those two properties, allows for a more reliable description of the chemical enrichment and star formation histories (SFH) of galaxies. 

In recent years, two major advancements in observational and broad-band SED fitting techniques have been achieved. First, the establishment of sophisticated SED modeling tools as the primary method of retrieving the main physical properties of galaxies \citep[see][]{Pacifici_2023ApJ...944..141P}. These SED fitting algorithms can combine panchromatic datasets from various observatories, while taking advantage of Bayesian statistics ({\tt MAGPHYS}, \citealt{da_Cunha_2008MNRAS.388.1595D}; {\tt CIGALE}, \citealt{Boquien_2019A&A...622A.103B}) and Monte Carlo sampling techniques ({\tt BEAGLE}, \citealt{Chevallard_2016MNRAS.462.1415C}; {\tt BAGPIPES}, \citealt{Carnall_2018MNRAS.480.4379C}; {\tt PROSPECT}, \citealt{Robotham_2020MNRAS.495..905R}; {\tt Prospector}, \citealt{Johnson_2021ApJS..254...22J}). The second advancement has to do with the use of wide-field cameras such as MegaCam \citep{Boulade_2003SPIE.4841...72B} and Hyper Suprime-Cam \citep{Miyazaki_2018PASJ...70S...1M, Aihara_2019PASJ...71..114A}. Wide-field cameras facilitate an important breakthrough on cosmological studies by providing the capability to scan several square degrees at a time, with high angular resolution and large survey speed. Collecting deep multi-wavelength data of different galaxy samples, at different cosmic epochs, is essential to get a better grasp on how galaxies form and evolve through cosmic time. 

A tremendous progress has been made in the field of panchromatic photometric surveys both at low redshift, for example, SDSS \citep{York_2000AJ....120.1579Y}, GAMA \citep{Driver_2011MNRAS.413..971D}, and high redshift, such as CANDELS \citep{Grogin_2011ApJS..197...35G, Koekemoer_2011ApJS..197...36K}, UltraVISTA \citep{Muzzin_2013ApJS..206....8M}, and HerMES \citep{Oliver_2012MNRAS.424.1614O}. These photometric surveys often include a great number of broad-band and narrow-band filters that guarantee a sufficient spectral coverage. Surveys such as the COSMOS \citep{Scoville_2007ApJS..172....1S, Weaver_2022ApJS..258...11W}, COMBO-17 \citep{Wolf_2008A&A...492..933W}, ALHAMBRA \citep{Moles_2008AJ....136.1325M}, SHARDS \citep{Perez_Gonzalez_2013ApJ...762...46P} or PAU/J-PAS \citep{Benitez_2009ApJ...691..241B, Abramo_2012MNRAS.423.3251A}, contain more than 20 broad, intermediate, and narrow wavebands that completely cover the optical and near-infrared (NIR) regime. The extensive spectral coverage of these surveys enables a better precision on the measurement of photometric redshifts, for much larger galaxy samples than would be possible using spectroscopy. In principle, fitting these multi-wavelength data with SPS models should also enable more reliable estimates on the stellar properties. Although this remains true for the stellar mass which can be estimated robustly (within 0.3~dex) from SED fitting \citep[e.g.][]{Muzzin_2009ApJ...701.1839M, Wuyts_2009ApJ...696..348W, Conroy_2009ApJ...699..486C, Pacifici_2023ApJ...944..141P}, large uncertainties persist on the recovery of stellar ages (either luminosity-weighted or mass-weighted ages), stellar metallicities, and consequently on the SFHs.  

The relative quantitative impact of these large uncertainties on the stellar properties is not fully understood. Such uncertainties may arise due to certain complex physical processes that are very difficult to disentangle from the photometric SEDs alone, such as the age-dust-metallicity degeneracies \citep[e.g.][]{Bell_2001ApJ...550..212B}. For example, the UV slope of a galaxy may appear red either due to the lack of star-formation activity or the attenuation of UV light by dust in the star-forming regions. Another known degeneracy exists between age and metallicity, where the effects of stellar age on optical colors is degenerate with changes in metallicity \citep[e.g.][]{Walcher_2011Ap&SS.331....1W}.

One way to mitigate these degeneracies is to perform a panchromatic SED fitting and taking advantage of the energy balance principle: all energy absorbed by dust in the rest-frame ultraviolet (UV) is re-radiated in the far-infrared (FIR). \citet{Haskell_2023MNRAS.525.1535H} showed evidence that the age-dust degeneracy is better constrained by including FIR data, with the energy balance SED modeling being effective up to $z\sim4$. Notwithstanding, FIR photometry at intermediate and high-redshifts is not as reliable as in the local Universe. \citet{Leja_2017ApJ...837..170L} showed that it is possible to mitigate the age-dust degeneracy, when fitting the UV-MIR photometry alone, by constraining the IR priors of the dust emission. Another way to break these degeneracies at intermediate and high-redshifts is to complement the photometric SEDs with spectroscopic data \citep[e.g.][]{Worthey_1994ApJS...94..687W, Bruzual_2003MNRAS.344.1000B, Trager_2000AJ....120..165T, Gallazzi_2009ApJS..185..253G}. The numerous absorption spectral features can help to constrain the SFH and chemical composition of galaxies. More recently, \citet{Tacchella_2022ApJ...926..134T} also highlighted the importance of spectroscopy to constrain the stellar metallicity and the necessity of both spectroscopy and photometry to alleviate the dust–age–metallicity degeneracy.

\begin{figure*}[t]
    \centering
    \includegraphics[width=18cm]{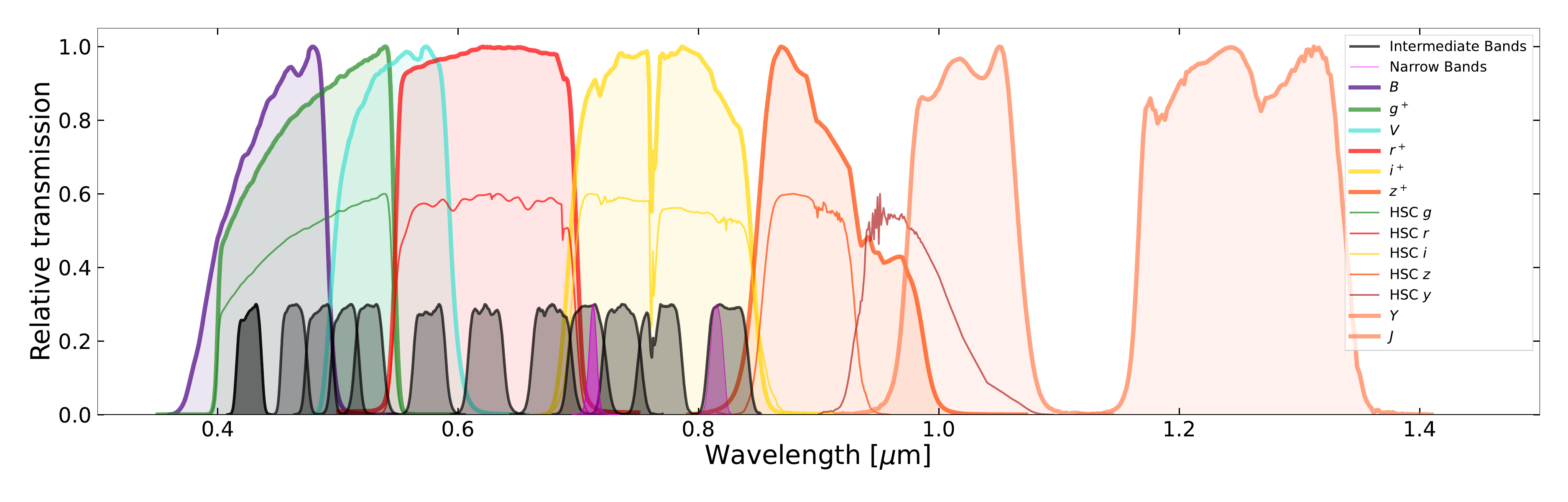}
    \caption{Relative transmission curves for the photometric bands used by the COSMOS survey. The magenta and black curves represent the narrow (NB711, NB816) and intermediate (IB427, IB464, IA484, IB505, IA527, IB574, IA624, IA679, IB709, IA738, IA767, IB827) Subaru Suprime-Cam bands, respectively. The remaining colored curves represent the optical broad bands ($B$, $g^+$, $V$, $r^+$, $i^+$, and $z^+$) from the Subaru Suprime-Cam and HSC, and the UltraVISTA Y, J broad bands in the NIR. The profiles are normalized to a peak transmission of 1.0 for the Suprime-Cam broad bands, to 0.6 for the HSC broad bands, and to 0.3 for the intermediate and narrow bands.}
    \label{fig:cosmos_2020_filter_set}
\end{figure*}
 
Despite the coarse spectral resolution of many photometric surveys, a recurring argument is that their spectral coverage is sufficient to resolve the stellar properties without the need of spectroscopic observations \citep{Pforr_2012MNRAS.422.3285P}. Yet, there has not been a quantitative comparison between the predicted spectra from SPS modeling of photometric SEDs and spectroscopic observations, for a statistically significant galaxy sample. In part, one challenge is that spectroscopic surveys cannot reach as deep as the photometric ones nor cover as large areas \citep[e.g.][]{Kriek_2011ApJ...743..168K, Bedregal_2013ApJ...778..126B, Whitaker_2013ApJ...770L..39W, Guzzo_2014A&A...566A.108G, Belli_2015ApJ...799..206B, Onodera_2015ApJ...808..161O, Kriek_2015ApJS..218...15K, Damjanov_2018ApJS..234...21D}. Of course some photometric surveys also accrued spectroscopic observations, for example, at low-redshift regimes there is SDSS and GAMA, while surveys at intermediate-redshift regimes include 3D-HST \citep{Brammer_2012ApJS..200...13B, Momcheva_2016ApJS..225...27M}, MOSDEF \citep{Kriek_2015ApJS..218...15K}, and VIPERS \citep{Scodeggio_2018A&A...609A..84S}. However, intermediate-redshift surveys usually trade-off signal-to-noise (S/N) with sample size, while focussing on bright emission lines originating from ionized gas in galaxies.  

The landscape in intermediate-redshift spectroscopy has changed with the Large Early Galaxy Astrophysics Census \citep[LEGA-C;][]{van_der_Wel_2016ApJS..223...29V, van_der_Wel_2021ApJS..256...44V} survey. The LEGA-C survey is an exceptional dataset that contains about 4,000 high S/N rest-frame optical spectra at redshift $0.6 \le z \le 1$ (or at lookback time of $\sim 7$~Gyr). The LEGA-C galaxy sample is $K_\mathrm{s}$-band selected and overlaps with the COSMOS field (see Section~\ref{sec:data}). The inclusion of optical spectra in the SED fitting can constrain the bulk formation age of the stellar populations, the metal enrichment history, and the burstiness of the SFH, through the fitting of key spectral features such as the Balmer lines (e.g. H$\delta$, H$\beta$, etc.) and various metal lines (e.g. Fe, Mg, etc.).

The scope of this paper is to test whether we can predict optical spectra from photometry with large wavelength baseline, without specific information on resolved spectral features. We argue that once the mean stellar age and metallicity are well constrained, so should the spectral indices. Hence, if the photometry fails to constrain the spectral indices, then it cannot produce good constraints on the stellar age and metallicity.

We will apply the SED fitting code \texttt{Prospector}\footnote{\url{https://github.com/bd-j/prospector}} \citep{Johnson_2021ApJS..254...22J} to the photometric catalog of COSMOS2020 \citep{Weaver_2022ApJS..258...11W}, in all available wavebands covering the rest-frame UV, optical, and NIR regimes. Then, we will compare the predicted model spectrum with the corresponding spectrum observed by the LEGA-C survey. We aim at quantifying the differences between model and observations: (i) by applying a $\chi^2$ test between the observed and predicted spectra, and (ii) by measuring the strength of two age- and metallicity-sensitive features, the \hda~and Fe4383. Typically, passive galaxies tend to be metal-rich (Fe4383~$>2$~\AA) with weak \hd~absorption (\hda~$<2$~\AA), whereas star-forming galaxies are usually metal-poor (low Fe4383) with a strong \hd~absorption line. A comparison of the output physical quantities is avoided here because the prior distributions of the free parameters in the SED modeling have a stronger impact on the inference of physical properties than on the predictions of directly observable quantities (i.e. the spectral indices).

This paper is structured as follows: in Section~\ref{sec:data} we describe the datasets we use and properties of the galaxy sample. In Section~\ref{sec:sed_fitting} we describe the SED fitting algorithm that we use to predict the model spectra. In Section~\ref{sec:results} we present the results of our analysis in a qualitative and quantitative manner. In Section~\ref{sec:discussion} we discuss the implications of our results, and finally in Section~\ref{sec:conclusions} we summarize our key findings and conclusions.

\section{Data and sample} \label{sec:data}

The LEGA-C sample \citep[DR3;][]{van_der_Wel_2021ApJS..256...44V} is selected based on the $K_\mathrm{s}$-band magnitude, taken from the Ultra Deep Survey with the VISTA telescope (UltraVISTA) catalog \citep{Muzzin_2013ApJS..206....8M, Muzzin_2013ApJ...777...18M}. LEGA-C contains 4081 spectra of 3741 unique galaxies (340 spectra are duplicate observations). For more details about the goals and design of the survey we refer the readers to \citet{van_der_Wel_2016ApJS..223...29V},   \citet{Straatman_2018ApJS..239...27S}, and \citet{van_der_Wel_2021ApJS..256...44V}.

The UltraVISTA catalog overlaps with the COSMOS field. Therefore, we are using the photometric data from the most recent COSMOS catalog \citep[COSMOS2020;][]{Weaver_2022ApJS..258...11W}. The first step in our analysis was to match the two catalogs. We used a rather conservative distance-separation of 0$^{\prime\prime}$.3 between the sky coordinates. After matching the two catalogs, we ended up with a sample of 3531 galaxies\footnote{If we use an even larger distance-separation of 1$^{\prime\prime}$, a match is returned for 3623 galaxies.}.

\subsection{Spectroscopic observations}

The spectroscopic observations of LEGA-C were carried out over the course of 4~years, using the now decommissioned VIMOS spectrograph \citep{Le_Fevre_2003SPIE.4841.1670L} at ESO's Very Large Telescope (VLT). The effective spectral resolution of LEGA-C is $R\sim3500$, with a typical observed wavelength range of 6300~\AA$~< \lambda <~$8800~\AA~or rest-frame $\sim$ 3000~\AA~$< \lambda <~$5550~\AA. The average S/N of the spectra in our sample is $\sim16$~\AA$^{-1}$. For the purpose of our analysis, we discarded galaxies with a S/N$<3$~\AA$^{-1}$. By applying this cutoff, 3217 galaxies remained in the sample.

The optical spectrum of a galaxy includes many absorption lines. From galaxy to galaxy, those absorption lines show small variations in terms of flux density and they are often pretty weak. Measuring the absorption line indices is a challenging effort that may suffer from various systematic effects in the sky subtraction, the noise model, and the wavelength calibration. Also, depending on whether or not the variance of the spectrum is considered, a bias can be introduced in the measurement of the equivalent width of the absorption line indices. In LEGA-C, extra care was given to reduce as much as possible those systematics and biases by employing an approximately bias-free method, described analytically in \citet{van_der_Wel_2021ApJS..256...44V}. A catalog was released with the Lick indices of 20 spectral absorption features, corrected for emission.

\subsection{Photometric observations}

The latest data release from the COSMOS survey \citep{Scoville_2007ApJS..172....1S} includes two multi-wavelength photometric catalogs that were obtained from two independent methodologies. The CLASSIC catalog uses a Point-spread function (PSF) homogenization and aperture-match photometry, while the FARMER catalog employed a model-based photometry that does not operate on PSF-homogenized images. The new catalogs gain almost one order of magnitude in photometric redshift precision and have deeper observations in the optical bands. For a detailed discussion on the photometric methods used to produce the data in the COSMOS2020 catalogs we refer to \citet{Weaver_2022ApJS..258...11W}.  

Here, we use the CLASSIC catalog for two reasons. Firstly, the UltraVISTA broad-band photometry \citep{Muzzin_2013ApJ...777...18M} was employed in LEGA-C to calibrate the galaxy spectra. Similar to the CLASSIC catalog, UltraVISTA contains PSF-matched photometry \citep{Muzzin_2013ApJ...777...18M}. Secondly, the FARMER catalog does not contain the Subaru Suprime-Cam broad-bands, which are included in the CLASSIC and UltraVISTA catalogs, as they suffer from high spatial PSF variability \citep{Weaver_2022ApJS..258...11W}. For simplicity, we will refer to the CLASSIC catalog as COSMOS2020.  

We use a collection of 27 photometric bands in the optical and near-infrared that cover the wavelength range of the LEGA-C spectroscopic data. Figure~\ref{fig:cosmos_2020_filter_set} displays the broad, intermediate, and narrow-bands that we use in our analysis. Basically, we work with the Subaru Suprime-Cam and Hyper-Suprime Cam (HSC) in the optical bands, and the UltraVISTA $Y, J$ bands in the near-infrared. There are two reasons why we do not use any photometric data beyond the $J$-band: (i) \citet{van_der_Wel_2021ApJS..256...44V} showed a systematic mismatch between a synthesized photometry and the UltraVISTA $H-K_\mathrm{s}$ color, even after a zero-point correction was applied to the UltraVISTA bands ($B$, $V$, $r$, $i$, $z$, $Y$, $J$, $H$, $K_\mathrm{s}$), and (ii) due to this mismatch in the $H-K_\mathrm{s}$ color, the LEGA-C spectra were calibrated using the $BVrizYJ$ filter set.

We also corrected the photometric measurements for galactic extinction. This was done by using the $E(B-V)$ values from the \citet{Schlafly_&_Finkbeiner_2011ApJ...737..103S} dust map and the \citet{Fitzpatrick_1999PASP..111...63F} attenuation law ($R_V = 3.1$).

\begin{figure}[t]
    \centering
    \includegraphics[width=\columnwidth]{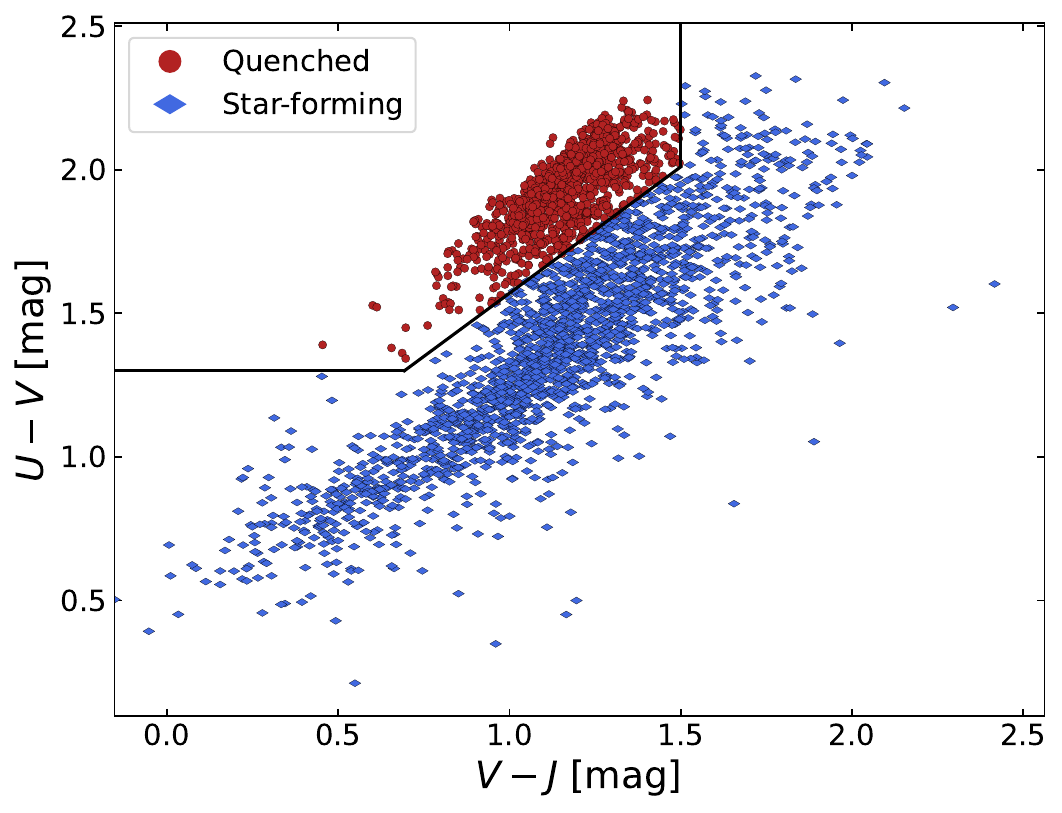}
    \caption{$UVJ$ diagram. We use the definition of \citet{Muzzin_2013ApJ...777...18M} (black line) to separate the galaxies in our sample into quiescent (red points) and star-forming (blue diamonds).}
    \label{fig:uvj_diagram}
\end{figure}

\begin{table*}
\caption{Free Parameters and their associated prior distribution functions in the {\tt Prospector} physical model.}
\begin{center}
\scalebox{1.}{
\begin{threeparttable}
\begin{tabular}{llr}
\hline 
\hline\\
Parameter & Description & Prior distribution \\
\hline\\
$\log\left(M_\star/\mathrm{M}_\odot\right)$ & total stellar mass formed & uniform: min=7, max=12\\[0.2cm]
$\log\left(Z_\star/\mathrm{Z}_\odot\right)$ & stellar metallicity & uniform: min=-1.98, max=0.4\\[0.2cm]
SFR ratios & Ratio of adjacent SFRs in the   & Student’s t-distribution with\\
           & eight-bin nonparametric SFH, & $\sigma=0.3$ and $\nu=2$\\
           & seven free parameters in total & \\[0.2cm]
$\log\left(Z_\mathrm{gas}/\mathrm{Z}_\odot\right)$ & gas-phase metallicity & uniform: min=-2, max=0.5\\
$z$ & redshift & fixed to LEGA-C spectroscopic redshift\\
\hline\\
$\hat{\tau}_{\lambda, 2}$ & diffuse dust optical depth & clipped normal: min=0, max=4, mean=0.3, $\sigma=1$\\[0.2cm]
$\hat{\tau}_{\lambda, 1}$ & birth-cloud dust optical depth & clipped normal in ($\hat{\tau}_{\lambda, 1}/\hat{\tau}_{\lambda, 2}$): min=0, max=4, mean=0.3, $\sigma=1$\\[0.2cm]
$n$ & slope of \citet{Kriek_2013ApJ...775L..16K} dust law & uniform: min=-1, max=0.4\\
\hline
\end{tabular}
\end{threeparttable}}
\label{tab:free_params_and_priors}
\end{center}
\end{table*}

Lastly, we compared the COSMOS2020 and UltraVISTA photometric catalogs by measuring the differences in terms of flux density. Specifically, we compared the flux densities in the photometric broad-bands: $B$, $V$, $r$, $i$, $z$, $Y$ and $J$. The typical differences in the subset ($BVrizYJ$) were below 0.1~dex, with only a few galaxies showcasing differences above 0.3~dex. We apply a third criterion in our sample, by discarding galaxies with flux residuals more than 0.3~dex in all photometric bands in the subset ($BVrizYJ$). The final galaxy sample contains 3130 galaxies in the redshift range of $0.6 < z < 1$. 

Figure~\ref{fig:uvj_diagram} depicts the $UVJ$ diagram of our sample. The rest-frame $U - V$ and $V - J$ colors were calculated by \citet{Straatman_2018ApJS..239...27S} through fitting template spectra to the UltraVISTA photometric SEDs. From the definition of \citet{Muzzin_2013ApJ...777...18M}, we separate galaxies into quiescent (red points) and star-forming (blue diamonds). Lastly, stellar mass estimates are available in the LEGA-C catalog \citep{van_der_Wel_2021ApJS..256...44V}. The stellar masses were estimated through SED fitting of the UltraVISTA broadband photometry \citep{Muzzin_2013ApJ...777...18M}. The stellar mass range of our final sample is $10^{8.9}$--$10^{12}$~M$_\odot$, with a mean value of $10^{10.8}$~M$_\odot$.

\begin{figure*}[!th]
    \centering
    \includegraphics[width=\textwidth]{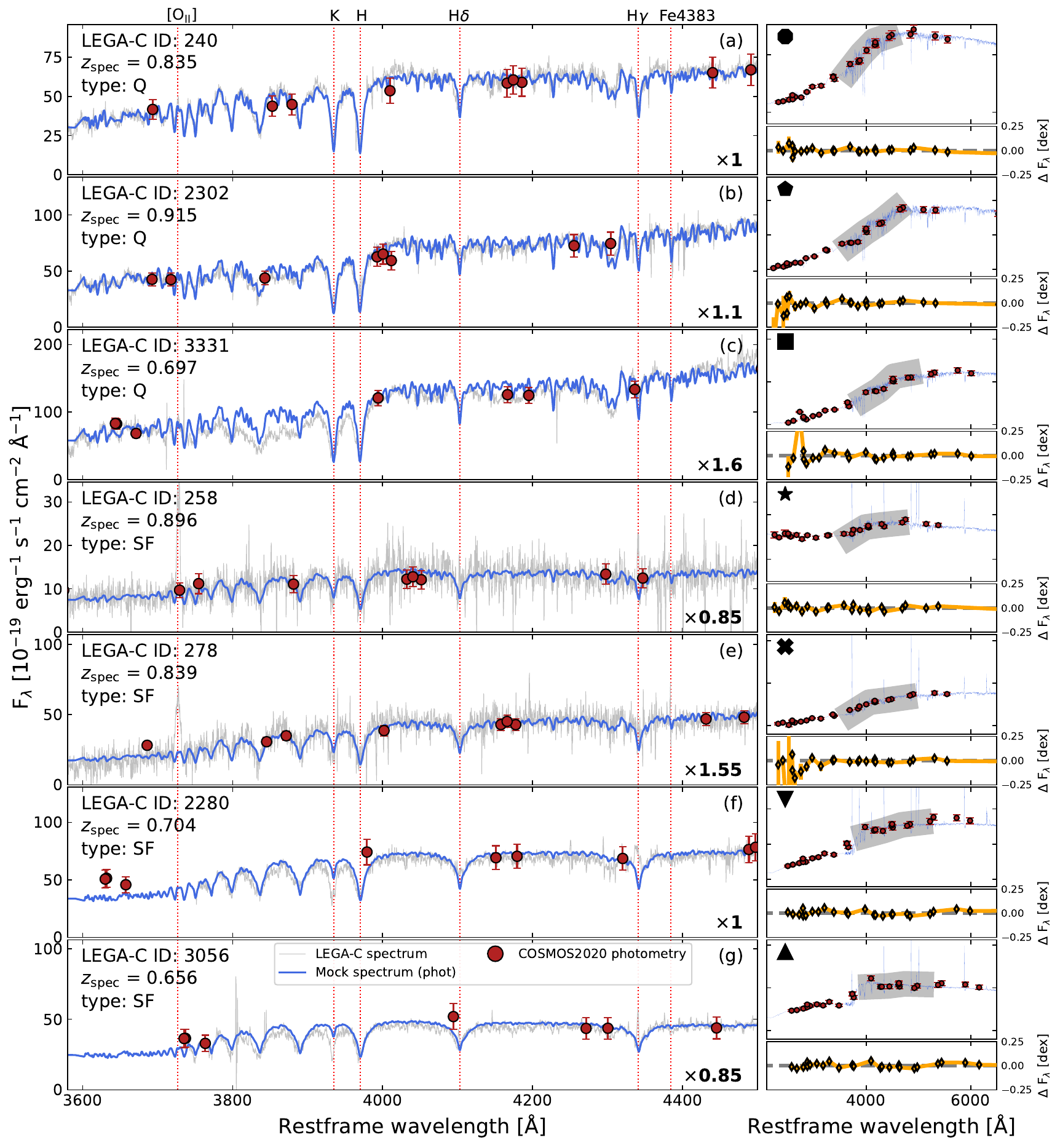}
    \caption{Seven randomly chosen examples of model spectra. Left column: The silver line represents the observed spectrum, and the blue line the model spectrum without the nebular emission, predicted by fitting the COSMOS2020 flux densities (red points). The dotted red lines indicate the central wavelength of various spectral absorption features. Panels (a)--(c) show spectra that belong to quenched galaxies, while the remaining panels display spectra that correspond to star-forming galaxies. For visualization purposes and when necessary, a multiplicative factor (bottom-right corner of each panel) was applied to match the photometry and predicted spectra to the observations. Right column: The top panel of each sub-figure shows the full COSMOS2020 observed SED (red points) and MAP model spectrum (blue line). The shaded silver region covers the wavelength range of the corresponding LEGA-C spectra. The bottom panel of each sub-figure shows the difference in dex between the observations and the best-fitting photometry.}
    \label{fig:sed_examples}
\end{figure*}

\section{SED Fitting} \label{sec:sed_fitting}

Fitting the photometric data is a computationally expensive process. As mentioned before, several SED fitting algorithms exist that utilize a Bayesian approach, to combine stellar, nebular, and dust models into composite stellar populations. In this paper, we use the {\tt Prospector} inference framework \citep{Johnson_2021ApJS..254...22J} to model the COSMOS2020 photometry. {\tt Prospector} adopts a Bayesian forward modeling and Monte Carlo sampling of the parameter space. This gridless `on-the-fly' modeling allows for a more complete exploration of the parameter space, compared to the early, grid-based SED fitting codes. {\tt Prospector} has been extensively tested in several studies by fitting both photometric and spectroscopic data, to retrieve various physical products. Applications include the SED modeling of nearby galaxies \citep{Leja_2017ApJ...837..170L, Leja_2018ApJ...854...62L} and high-redshift galaxies \citep{Leja_2019ApJ...877..140L}, dwarf galaxies \citep{Greco_2018ApJ...866..112G, Pandya_2018ApJ...858...29P}, as well as the retrieval of SFHs \citep{Leja_2019ApJ...877..140L} and dust attenuation properties \citep{Nagaraj_2022ApJ...932...54N} in galaxies.  

We created a model with 13 free parameters. The functions and range of the prior distributions are given in Table~\ref{tab:free_params_and_priors}, following the model presented in \citet{Leja_2019ApJ...876....3L}. We fixed the redshift to the robustly measured LEGA-C spectroscopic values\footnote{The spectroscopic redshifts are in an exceptional agreement with the photometric ones from the COSMOS2020 catalog.}. One advantage of {\tt Prospector} is the availability of nonparametric SFHs. In our analysis we employed the `continuity' SFH with a Student's-t prior distribution described thoroughly in \citet{Leja_2019ApJ...876....3L}. This particular prior favors a smooth SFH without sharp transitions in SFR(t), according to the regularization schemes by \citet{Ocvirk_2006MNRAS.365...46O} and \citet{Tojeiro_2007MNRAS.381.1252T}. We use eight time elements in the nonparametric SFH model, specified in lookback time. The first two time bins are fixed at 0--30~Myr and 30--100~Myr to capture recent variations in the SFH of galaxies. To model the oldest stellar population in a particular galaxy, a third time bin is placed at ($0.85~t_\mathrm{univ} - t_\mathrm{univ}$), where $t_\mathrm{univ}$ is the age of the Universe at the observed redshift. The remaining five bins are spaced equally in logarithmic time between 100~Myr and $0.85t_\mathrm{univ}$. 

The stellar metallicity is also a free parameter with a flat prior. Assuming a constant stellar metallicty history can have an impact on the derived physical properties. For example, \citet{Thorne_2022MNRAS.517.6035T} fitted the photometric SEDs of 7000 low-redshift GAMA galaxies and demonstrated that there are severe systematic offsets in the recovered stellar ages depending on the assumed metallicity prescriptions. \citet{Thorne_2022MNRAS.517.6035T} found that using a fixed metallicity for all galaxies leads to systematic offsets of 0.5~dex at intermediate ages. On the other hand, a constant metallicity history (like the one we use in this paper) can underestimate the older ages up to 0.1~dex, as opposed to an evolving metallicity history, due to the age-metallicity degeneracy. Moreover, in an upcoming study by \citet{Gallazzi_2023ApJ}, the authors fit the Lick indices of the LEGA-C spectra by assuming either a fixed or a variable mass-weighted metallicity and find no significant offsets in the stellar population ages. Therefore, assuming a constant metallicty history does not have a strong effect on the derived physical properties.

{\tt Prospector} utilizes the Flexible Stellar Populations Synthesis ({\tt FSPS}) stellar populations code \citep{Conroy_2009ApJ...699..486C}, to model the stellar properties. We adopted the default SPS parameters in {\tt FSPS}, that is the MILES stellar library and the MIST isochrones. We chose the \citet{Chabrier_2003PASP..115..763C} initial mass function (IMF) in our modeling. The nebular continuum and line emission is generated through a grid of models \citep{Byler_2017ApJ...840...44B} that were produced with \texttt{CLOUDY} \citep{Ferland_2013RMxAA..49..137F}. A flat prior was given for the gas-phase metallicity. The remaining free parameters are related to a variable dust attenuation law \citep{Kriek_2013ApJ...775L..16K}, which are also given a flat prior distribution. Lastly, we used the nested sampler {\tt dynesty} \citep{Speagle_2020MNRAS.493.3132S}, that simultaneously estimates both the Bayesian evidence and the posterior distributions, while it allows a dynamic sampling of the parameter space to maximize a chosen objective function as the fit proceeds. Out of the 3130 galaxies in our sample, {\tt Prospector} converged to a solution for 3101 galaxies.

We note here that after we fit the COSMOS2020 photometry, we exclude the nebular emission from our maximum a posteriori (MAP) model SED as we are only interested in the absorption lines of the predicted spectrum.   

\section{Results} \label{sec:results}

In this section, we evaluate the results of fitting the COSMOS2020 photometry with {\tt Prospector}. In Section~\ref{subsec:obs_vs_mod_spec} we present a qualitative comparison between the observed LEGA-C spectra and the model spectra predicted with SED fitting. In Section~\ref{subsec:spectral_comp} we provide a more quantitative comparison of the observed vs predicted spectra by measuring the Lick indices \citep{Worthey_1994ApJS...94..687W} of two key absorption spectral features: \hda~and Fe4383.  

\subsection{Observed vs model spectra} \label{subsec:obs_vs_mod_spec}

We retrieve 3101 model spectra by fitting the COSMOS2020 photometry with {\tt Prospector}. Those spectra represent the MAP SEDs\footnote{No broadening due to velocity dispersion was applied to the model spectra. The model spectra are at their original resolution of the MILES  stellar library, that is 2.3~\AA~[full width at half-maximum (FWHM)].}. Figure~\ref{fig:sed_examples} shows seven randomly selected examples of model spectra predicted with {\tt Prospector}, for both quiescent and star-forming galaxies. For visualization purposes, the COSMOS2020 photometry and predicted spectra in Fig.~\ref{fig:sed_examples} have been re-scaled with a multiplicative factor, which does not affect the spectral feature strength, to match the observed spectra. This multiplicative factor is the median of the ratio of the two spectra across the rest-frame wavelength range $\sim$ 3000~\AA~$< \lambda <~$5550~\AA. In Table~\ref{tab:example_params}, we provide the physical estimates of those seven galaxies from the SED fitting with {\tt Prospector}, in order to get a more precise sense of their physical properties.

\begin{table}
\renewcommand{\arraystretch}{2.0}
\caption{Best-fit values and associated 1--$\sigma$ error for the seven randomly chosen galaxies shown in Fig.~\ref{fig:sed_examples}. The physical properties were retrieved from the SED fitting of COSMOS2020 photometry with {\tt Prospector}.}
\begin{center}
\scalebox{1.}{
\begin{threeparttable}
\begin{tabular}{lcccc}
\hline 
\hline
LEGA-C ID & $\log M_\star$ & $\log Z_\star$ & mw-Age & SFR\\
     & [M$_\odot$] & [Z$_\odot$] & [Gyr] & [M$_\odot$/yr] \\
\hline
240  & $11.04_{0.03}^{0.05}$ & $-1.21_{0.2}^{0.16}$ & $9.7_{0.6}^{2.6}$ & $0.01_{0.01}^{0.29}$\\
2302 & $11.40_{0.05}^{0.06}$ & $-0.58_{0.13}^{0.18}$ & $12.9_{0.6}^{0.1}$ & $0.04_{0.04}^{0.35}$\\
3331 & $11.24_{0.04}^{0.06}$ & $-1.04_{0.16}^{0.2}$ & $9.5_{0.7}^{0.8}$ & $0.07_{0.07}^{0.58}$\\
258  & $10.25_{0.1}^{0.06}$ & $-0.34_{0.28}^{0.24}$ & $7.5_{0.8}^{0.6}$ & $4.43_{1.82}^{2.78}$\\
278  & $10.72_{0.1}^{0.14}$ & $-0.29_{0.45}^{0.6}$ & $8_{1.6}^{4.9}$ & $6.03_{3.76}^{11.35}$\\
2280 & $10.86_{0.1}^{0.07}$ & $-1.38_{0.21}^{0.36}$ & $3.4_{1.4}^{1.6}$ & $265_{237}^{172}$\\
3056 & $10.6_{0.07}^{0.05}$ & $-1.77_{0.28}^{0.37}$ & $6.4_{0.6}^{0.4}$ & $20.4_{13.4}^{10.6}$\\
\hline
\end{tabular}
\end{threeparttable}}
\label{tab:example_params}
\end{center}
\end{table}

\begin{figure}[t]
\centering
\includegraphics[width=\columnwidth]{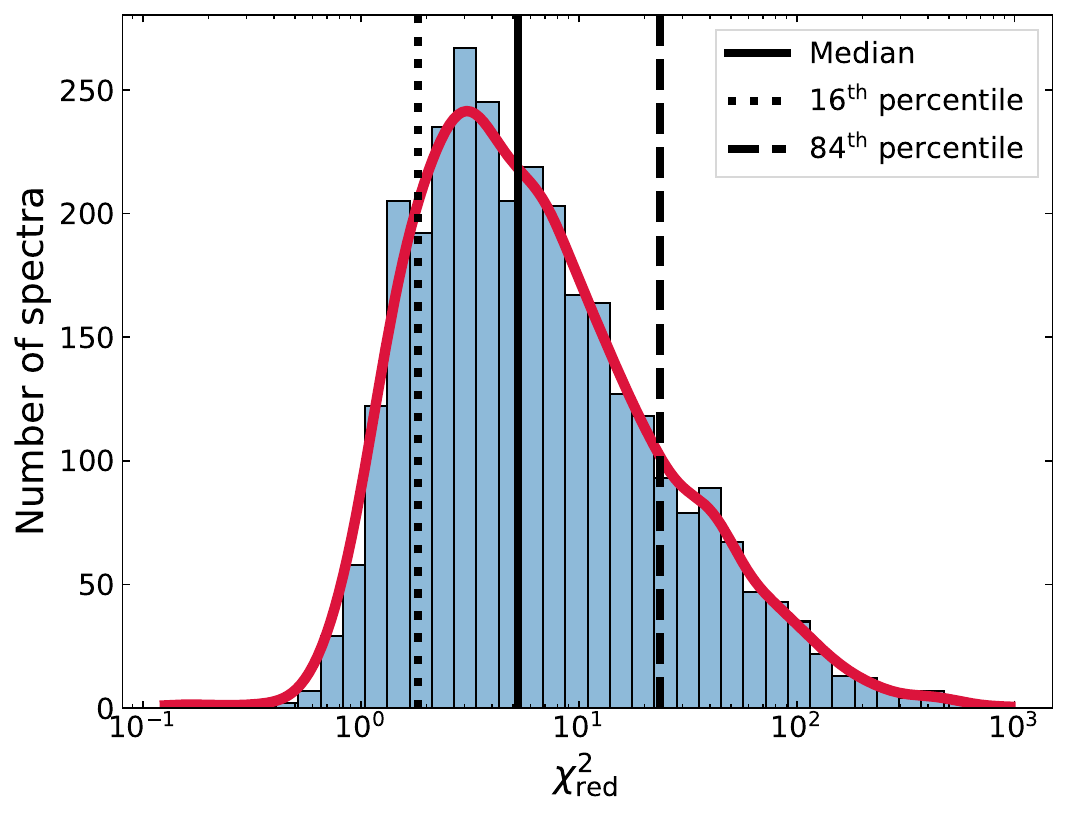}
\caption{Distribution of the reduced $\chi^2$ between the predicted spectra with {\tt Prospector} and the observed LEGA-C spectra. The Kernel Density Estimate (KDE) distribution is shown in red, while the solid black line shows the median value. The dotted and dashed black lines indicate the $16^\mathrm{th}$ and $84^\mathrm{th}$ percentiles of the distribution, respectively.}
\label{fig:chisqr}
\end{figure}

\begin{figure*}[t]
    \centering
    \includegraphics[width=18cm]{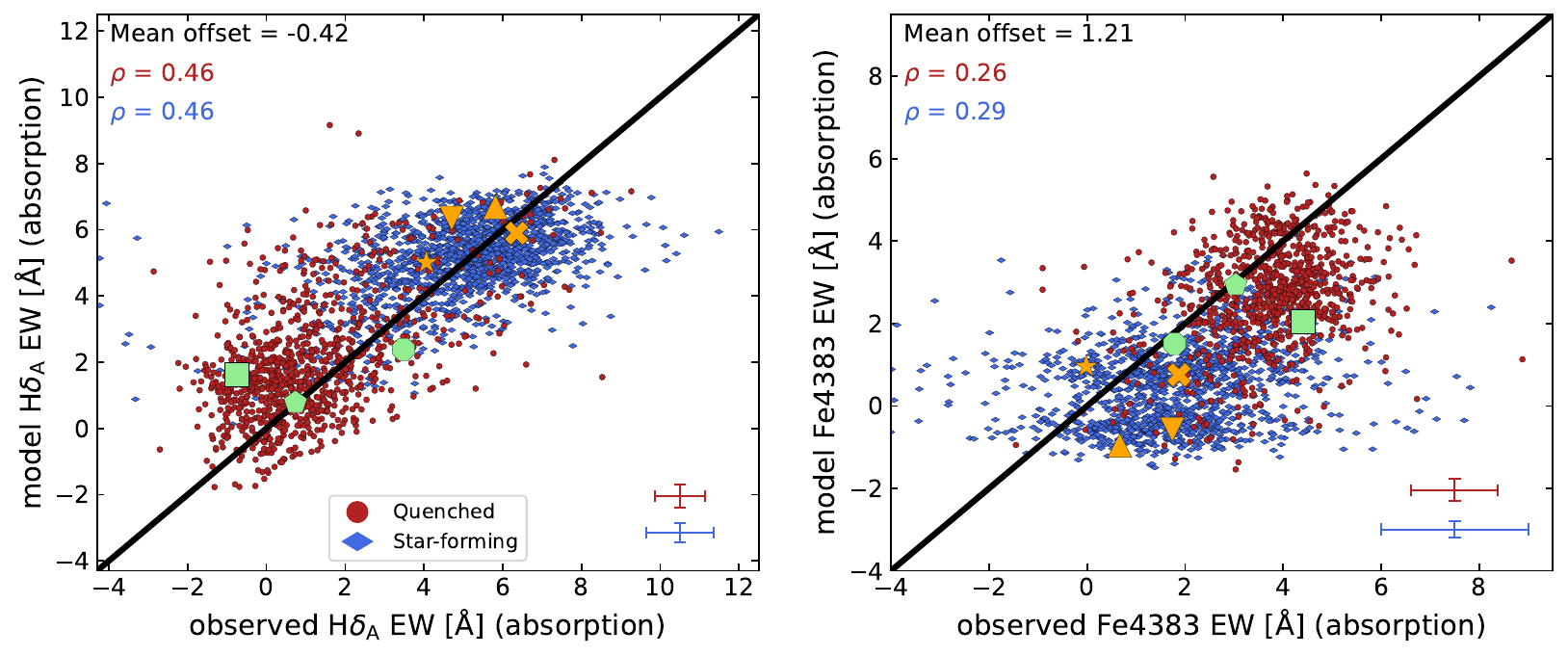}
    \caption{Predicted \hda~and Fe4383 absorption features from {\tt Prospector} fits to the COSMOS2020 photometry compared to observations. Galaxies are color-coded by their $UVJ$--diagram classification to star-forming (blue diamonds) and quenched (red points). The absorption-only models are compared to the observed values (corrected for emission). The black line shows the one-to-one relation. The various markers correspond to the star-forming (orange) and quenched (green) galaxies shown in Fig.~\ref{fig:sed_examples}. The statistics of the mean offset and the Spearman's rank correlation coefficient ($\rho$), are shown in the top-left corner of each panel.} The median uncertainties of each axis and galaxy population are shown in the bottom-right corner of each panel.
    \label{fig:hd_fe4383_obs_vs_prd}
\end{figure*}

Qualitatively, it is striking how well the shape of the spectrum and its spectral features can be retrieved by just fitting the broad-band and narrow-band photometry. Some cases (panels a and f) are even near perfect, while others show systematic differences between predicted and observed spectra (panels d and e). There are also cases where the predicted spectrum follows the spectral shape closely yet there is a wavelength-dependent offset (panels c and g) suggesting that the global spectral shape of the LEGA-C spectrum and COSMOS2020 are inconsistent. Regarding the LEGA-C galaxies 2280 and 3056, a slight offset can be seen in the absorption line wavelengths of the observed spectrum and the model. However, these apparent offsets are due to a blueshifted \hg~emission (in the case of 2280) or a line-strength dependent wavelength due to blending of multiple lines (e.g., the 3933~\AA~line). 

To explore the overall quality of the predicted spectra to the observations, we examine the distribution of the reduced $\chi^2$ values:

\begin{equation} \label{eq:chi2_red}
\\\\\ \chi^2_\mathrm{red} = \frac{1}{\nu} \sum_{i=1} \frac{\left(O_i - P_i\right)^2}{\sigma_i^2},
\end{equation}

\noindent where $O_i$ are the observed spectra, $P_i$ are the model spectra, $\sigma_i$ are the observed uncertainties, and $\nu$ is the number of wavelength elements minus the number of free parameters. The $\chi^2_\mathrm{red}$ distribution is shown in Fig.~\ref{fig:chisqr}. We find that the peak of the $\chi^2_\mathrm{red}$ distribution is at $\sim3.1$, while the median value of the histogram is skewed to a higher value (5.3). Out of the 3101 predicted spectra, only the 18\% has a $\chi^2_\mathrm{red}\le2$. About 68\% of the predicted spectra has a $\chi^2_\mathrm{red}\ge3$. This is indicative of the overall disagreement between the observations and the predicted spectra from photometry.

To further quantify how well we can predict the spectra from SED fitting, we measured the Lick indices of two key absorption lines \hda~and Fe4383. These spectral features can be used as proxies of the age and metallicity, respectively. The results are shown in the following section.  

\subsection{Spectral Measurements} \label{subsec:spectral_comp}

In our analysis, we are using the accurately measured Lick indices from \citet{van_der_Wel_2021ApJS..256...44V}. As for the predicted absorption spectra from {\tt Prospector}, we measure the same 20 Lick indices\footnote{Same as those provided in the LEGA-C catalog by \citet{van_der_Wel_2021ApJS..256...44V}.} with the {\tt python} package {\tt pyphot}\footnote{\url{https://github.com/mfouesneau/pyphot}}. Our choice to use the {\tt pyphot} package is based on the fact that the model spectra do not have noise. On the other hand, running the {\tt pyphot} algorithm on the observed spectra would have led to biases in the index measurements (due to asymmetries induced by strongly wavelength dependent noise), and thus should be avoided. In any case, to test the reliability of {\tt pyphot} we measured the lick indices of synthetic data with known absorption line index values, generated with the MILES SPS library. The {\tt pyphot} package was able to retrieve the true values of the Lick indices of the synthetic data with an excellent accuracy ($\rho = 1$).

Figure~\ref{fig:hd_fe4383_obs_vs_prd} shows a comparison between the observed and predicted Lick indices. Specifically, we compare the age- and metallicity-sensitive \hda~and Fe4383 spectral features. The predicted values of the absorption lines are derived from the SED at the MAP probability. The corresponding 16--84 percentile uncertainties are estimated by drawing 500 SEDs weighted by the {\tt dynesty} weights, and measuring the values of the absorption lines from those 500 SEDs.

For the sample as a whole there is a strong correlation $\rho = 0.75$ between the predicted and observed \hda~absorption line strength (Fig.~\ref{fig:hd_fe4383_obs_vs_prd}), with no significant systematic offset (0.42~\AA). From this plot we see a clear separation between the passive and the star-forming galaxies. Quiescent galaxies are characterized by weak \hd~absorption (\hda~$\sim 0.8 $~\AA), while star-forming galaxies show strong \hd~absorption (\hda~$>4.9$~\AA). The bimodality seen in the observed values \citep[also see][]{Straatman_2018ApJS..239...27S, Wu_2018ApJ...855...85W} is reproduced in the distribution of predicted \hda~line strengths.

For quiescent and star-forming galaxies separately the correlation is, naturally, weaker ($\rho = 0.46$ in both cases). For quiescent galaxies the correlation is driven by a tail of young post-starburst galaxies with strong \hda~lines. For star-forming galaxies there is a non-unity slope in the distribution, which reflects either that high-\hda~galaxies have underestimated predicted \hda~values (and \textit{vice versa}) or that the relatively large uncertainties on the weak \hda~lines in the LEGA-C spectra introduces scatter. We examine the latter option by taking the predicted values as ground truth and perturbing those by the LEGA-C measurement uncertainties to induce scatter. The resulting scatter is 1.69~\AA, which is smaller than the observed scatter of 1.96~\AA~in Fig.~\ref{fig:hd_fe4383_obs_vs_prd}. Whereas there is no systematic offset for star-forming galaxies, quiescent galaxies show a strong systematic offset of $\sim 0.85$~\AA, which is reminiscent of the offset between simulated synthetic spectra and LEGA-C spectra analyzed by \citet{Wu_2021AJ....162..201W}. 

In the right-hand panel of Fig.~\ref{fig:hd_fe4383_obs_vs_prd}, we compare the predicted and observed Fe4383 feature strength. Again, the galaxy bimodality seen in the observed values is reproduced in the distribution of the predicted Fe4383~line strengths. Quiescent galaxies show strong Fe4383 absorption (Fe4383~$\sim 3.75 $~\AA), while star-forming galaxies show weak Fe4383 absorption (\hda~$\sim 1.9 $~\AA). While there is an overall correlation, there is more scatter compared to \hda, as well as a substantial systematic offset (1.21~\AA). For quiescent and star-forming galaxies separately, there is only weak correlation. Furthermore, the predicted Fe4383 values of the star-forming galaxies seem to be stagnated around two particular values. This is due to the limited range of metallicities and element abundance ratios in the current SPS models that we are using, resulting in a limited variety of absorption features. We find similar systematic offsets for all measured indices from the model spectra (see Appendix~\ref{ap:A}).

In Fig.~\ref{fig:hd_fe4383_obs_vs_prd} we also indicate the measured Lick indices of the seven randomly selected galaxies from Fig.~\ref{fig:sed_examples}. For the galaxies that there is a good agreement between the observed and model spectrum, such as panels (a) and (b), we also notice a very good agreement on their respective measured indices. The differences in the Lick indices of galaxies increase as the model spectrum deviates more and more from the observed one, for instance the galaxies in panels (c) and (g).  

Finally, we note that the uncertainties on the predicted feature strengths are much smaller than the LEGA-C measurement uncertainties. This implies that either the formal uncertainties in the model spectra are underestimated, or that 20-hour spectra of $z\sim 1$ galaxies is insufficient to match the information content of 27-band photometry from the UV to the near-IR.

\section{Discussion} \label{sec:discussion}

The inference of physical properties from data involves many steps, each of which introduce a new level of uncertainty. In this paper, we examined to what extent photometry can be used to predict spectra, which addresses the uncertainty due to the loss of spectral information between spectroscopy and photometry. However, other uncertainties are also present and must be evaluated too. These roughly fall into two categories: uncertainties on the data level (Sec.~\ref{subsec:data_level}), and uncertainties on the interpretation level (Sec.~\ref{subsec:interpretation_level}). 

\subsection{Predicting spectra from photometry} \label{subsec:data_level}

In the previous section we showed that the observed and predicted spectral indices agree to a certain degree qualitatively and quantitatively. Nevertheless, the offsets seen in both \hda~and Fe4383, are large enough to have direct implications on the resulting stellar ages (either luminosity-weighted or mass-weighted ages) and metallicity. Consequently, the discrepancies found between the measured and predicted spectral features will also have a strong impact on the derived SFHs.

\begin{figure*}[t]
    \centering
    \includegraphics[width=\textwidth]{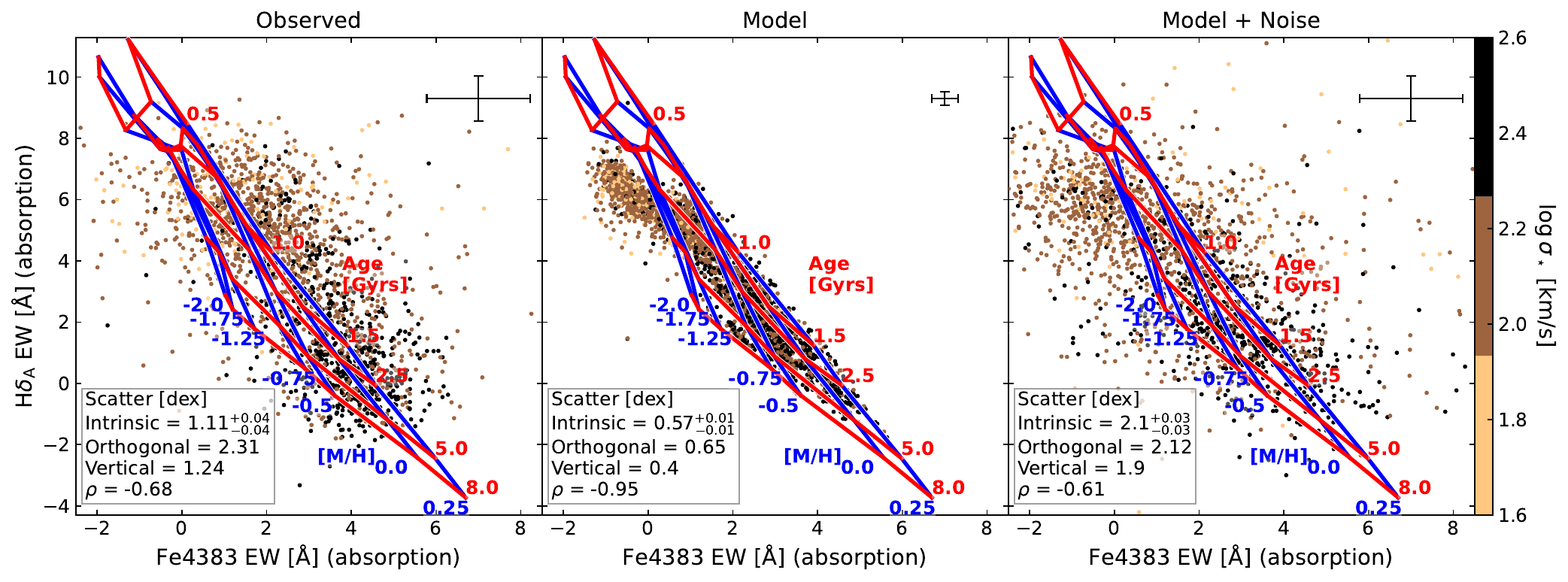}
    \caption{Left: Observed \hda~versus Fe4383. Middle: Predicted \hda~versus Fe4383. Right: Perturbed model values according to the individual observed errors. Galaxies are color-coded with $\log \sigma_\star$ from the LEGA-C DR3 catalog. The SSP model grid is shown with the red (Age) and blue (metallicity) lines. The statistics of the scatter and the Spearman's rank correlation coefficient ($\rho$) are shown in the bottom-left corner of each panel. The median uncertainties of each axis are shown in the top-right corner of each panel.}
    \label{fig:hd_vs_fe4383}
\end{figure*}

In Fig.~\ref{fig:hd_vs_fe4383}, we show the relation between the two aforementioned spectral features. On the left-hand panel of the figure we show the observed relation and on the middle panel we show the predicted relation. We also show the line strengths of the simple stellar population (SSP) model grid. A third dimension is added to this figure by color-coding the points with the measured stellar velocity dispersion ($\sigma_\star$) from the observed spectra. We note that the trend with $\sigma_\star$ is similar in both relations. Galaxies with low \hda~and high Fe4383 values also have high velocity dispersion. Conversely, galaxies tend to have low $\sigma_\star$ for high \hda~and low Fe4383 values. Overall, this trend with $\sigma_\star$ is in agreement with the general picture that we know about galaxies \citep[e.g.][]{Wake_2012ApJ...751L..44W, McDermid_2015MNRAS.448.3484M, Straatman_2018ApJS..239...27S, Chauke_2018ApJ...861...13C}. High-$\sigma_\star$ galaxies ($\sigma_\star\ge170$~km/s) are usually older (\hda~$<2$~\AA) and metal-rich (Fe4383~$>2$~\AA), whereas galaxies with low $\sigma_\star$ are usually young, star-forming galaxies (high \hda) and metal-poor (low Fe4383). However, this does not mean that fitting the photometric SEDs would yield an accurate measurement of the stellar ages and metallicities. This is more clear when we look at the statistics of the relation. 

We notice that the dynamical range of the observed relation is moderately larger than the one predicted with SED fitting. The limited dynamical range of \hda~may be related to the use of the continuity SFH, which is smoothing out any bursts of star formation that would have allowed \hda~to take larger range of values. While using a more bursty SFH prior \citep[e.g.][]{Suess_2022ApJ...935..146S} would help reproduce the properties of some galaxies, this may also introduce spurious bursts for the bulk of the (non star-burst) population. Furthermore, by measuring the orthogonal and vertical scatter of the relation, we find that all values are significantly lower for the predicted relation. The reduced scatter in the predicted relation is unsurprising considering that the model spectra are free of noise. If we perturb the model values according to the individual observed uncertainties (see right-hand panel of Fig.~\ref{fig:hd_vs_fe4383}), then we immediately notice that the scatter around the relation becomes similar to the observed one.

In addition, the SSP model grid in Fig.~\ref{fig:hd_vs_fe4383} hints at possible limitations of the current SSP templates (e.g. stellar libraries, modeling of stellar evolutionary phases) to capture some of the variance in the observed spectra of galaxies, either due to incomplete stellar libraries or poorly calibrated physics \citep{Conroy_2013ARA&A..51..393C}. Another type of limitation is the variability of the metal enrichment history. $\alpha$-enhancement and in general variable element abundance ratios may lead to inconsistencies and a poor match of the absorption features. Of course, these limitations would affect both photometric SED and spectral fitting. In any case, if the underlying model grid does not cover the observed range of properties then the spectra cannot be faithfully reproduced.

We fit the relation with a Bayesian fit weighted by the data uncertainties \citep{Lelli_2019MNRAS.484.3267L}, and we find that the median slope of the observed relation ($-1.583\pm0.001$) is steeper than the predicted one ($-1.334\pm0.007$). This means that young star-forming galaxies might appear to have lower metallicities and older ages than what the observed features suggest. As expected, fitting only the photometric SEDs can result in severe systematic offsets in the physical estimates, especially for the stellar metallicity estimates. Only when photometry is combined with spectroscopy it is possible to reduce the systematics in the derived physical properties of galaxies \citep[see][]{Johnson_2021ApJS..254...22J, Tacchella_2022ApJ...926..134T}. 

The systematic uncertainties in the photometry could be held partially responsible for the strong offsets in the derived stellar properties. To evaluate any possible biases in the photometry or in the SED fitting method, we performed a mock analysis by perturbing the original fluxes of the COSMOS2020 catalog within their corresponding uncertainties. Then, we fitted the mock observations with {\tt Prospector} and measured the Lick indices of the \hda~and the Fe4383, finding no significant changes in our original results (see Appendix~\ref{ap:B}). 

On the other hand, we find that the COSMOS2020 colors have systematically larger $B-V$ values (0.085~mag) and lower $V-i^+$ values (0.168~mag) than the corresponding UltraVISTA colors. The detected offsets in the optical colors certainly signal some level of inconsistency between the observed and predicted spectra. One possible explanation for such an offset could be differences in the zero-points. \citet{van_der_Wel_2021ApJS..256...44V} applied a zero-point correction to the UltraVISTA photometry so that the flux densities are independent of stellar population synthesis models. However, a comparison of the COSMOS2020 with the original UltraVISTA photometry \citep{Muzzin_2013ApJ...777...18M} also revealed similar offsets. Hence, the most likely explanation for these offsets may be due to subtle differences in methodology when performing the aperture-match photometry. Regardless, we should mention here that it is beyond the scope of this paper to apply or suggest any corrections to the COSMOS2020 catalog nor to investigate the origin of the offsets in the broad-band colors. We simply want to test whether all of the SPS information is included in a galaxy's SED or whether the optical spectra provide additional information. 

As mentioned in Sec.~\ref{subsec:spectral_comp}, another source of discrepancy is the systematic and random uncertainties in the spectroscopic index measurements from observations. The data reduction step could introduce additional systematics or bias the index measurements. For example, how someone deals with the sky subtraction and flux calibration of the observed spectra, could potentially have a systematic effect on the spectral indices, on a galaxy-by-galaxy basis, increasing the random uncertainty. In the case of LEGA-C, a bias-free approach was employed to measure the spectral indices, that suffers less from the varying noise of the wavelength elements \citep[see][for more details]{van_der_Wel_2021ApJS..256...44V}. We estimate that 13\% of the variance in the left-hand panel of Fig.~\ref{fig:hd_fe4383_obs_vs_prd} is due to random uncertainties in the spectral index measurements.

\subsection{Inferring ages and metallicity from photometry} \label{subsec:interpretation_level}

We fitted both broad-band and narrow-band photometry, and performed a comparison in data space, highlighting the differences between predicted and observed spectra, in terms of Lick index measurements. In a similar study, \citet{Wu_2021AJ....162..201W} also reported a spectral mismatch between LEGA-C and synthetic spectra generated from the IllustrisTNG TNG100 simulation \citep{Marinacci_2018MNRAS.480.5113M, Naiman_2018MNRAS.477.1206N, Nelson_2018MNRAS.475..624N, Pillepich_2018MNRAS.473.4077P, Springel_2018MNRAS.475..676S}. The cause of this mismatch could be either due to a difference in galaxy evolution physics or due to systematic uncertainties in the stellar population models. With some broader assumptions on the SFH, maybe it is possible to cover the space of observed indices. But, even if we assume that there are no errors in the data, and the model grid covers the full observed space, there is still an imperfect mapping from data to physical properties. 

The results of our analysis hint that the SFHs retrieved with photometric SED fitting do not capture the full complexity and dynamic range of real SFHs, hence failing to predict the detailed absorption features which contain additional information about the age distribution within galaxies and elemental abundances. Also, more systematic errors can be introduced when modeling the SED of a galaxy. For example, \citet{van_der_Wel_2021ApJS..256...44V} showed that the current models do not fit the rest-frame photometry beyond 1~$\mu$m, leading to systematic errors up to $\sim 20\%$ (see their Appendix~B). These errors ultimately propagate in the derived physical properties such as the stellar mass, star-formation rate, and other parameters that are inferred from SED fitting. 

Other related studies, choose to compare the derived parameters from SED fitting by including or not optical spectroscopy. For instance, \citet{Tacchella_2022ApJ...926..134T} fitted the UV-IR photometry for a sample of massive quiescent galaxies, which lie in the CANDELS survey footprint, with and without a spectrum (see their Appendix~B). \citet{Tacchella_2022ApJ...926..134T} showed that differences arise in the derived properties when fitting only photometry, only spectroscopy, and both photometry and spectroscopy together. They concluded that combining photometry and spectroscopy significantly improves the derivation of parameters, especially the stellar metallicity estimates. \citet{Webb_2020MNRAS.498.5317W} argued that a mass-metallicity prior is needed to constrain the stellar metallicity while fitting spectra, but even then large uncertainties persist (see their Appendix~B). 

A large number of broad-band and narrow-band filters certainly help to constrain the shape of a galaxy's SED, yet the retrieval of the stellar properties come with large systematics and uncertainties. That is why high-quality spectra are so important. High S/N and high-resolution spectroscopic data, such as those acquired by LEGA-C, are necessary to constrain the different spectral features when performing SED modeling. Notwithstanding, the use of spectra is not a panacea. It has been shown that different codes produce different estimates of the stellar properties \citep[e.g.][]{Pacifici_2023ApJ...944..141P}, even when using the same high-quality spectra and photometry \citep[e.g.][]{Kaushal_2023arXiv230703725K, Gallazzi_2023ApJ}. 

Informing our SED physical models with better motivated age and metallicity priors, and most importantly conditioning on the observed spectroscopic features, is absolutely necessary if we want to reduce the uncertainties and systematics when measuring the stellar properties of galaxies.  

\section{Summary \& Conclusions} \label{sec:conclusions}

We have predicted the \hda~and Fe4383 spectral features of the COSMOS field using the COSMOS2020 photometric catalog \citep{Weaver_2022ApJS..258...11W} and the SED fitting code {\tt Prospector} \citep{Johnson_2021ApJS..254...22J}. Modeling the broad-band and narrow-band photometry of galaxies at different cosmic epochs is a commonly used method for estimating the intrinsic physical properties of the unresolved stellar populations. Yet, the derived stellar properties come with large uncertainties. These uncertainties arise from the fact that only photometric SEDs cannot resolve the various spectral features that could potentially constrain the age and metallicity of stellar populations. 

Here, we compared the predicted values with their observed counterparts from the LEGA-C spectroscopic survey. We highlighted the differences between predictions and observations by presenting two key spectral absorption features, \hda~and Fe4383. While the global bimodality of star-forming and quiescent galaxies in photometric space is recovered with the model spectra, there is little to no correlation between the predicted and observed spectral indices within these sub-populations.

For now we caution that photometry-based estimates of stellar population properties are determined mostly by the modeling approach and not the physical properties of galaxies, even when using the highest-quality photometric datasets and state-of-the-art fitting techniques. When exploring new physical parameter space (i.e. redshift or galaxy mass) high-quality spectroscopy is always needed to inform the analysis of photometry.

\begin{acknowledgements}
We thank the anonymous referee for the valuable remarks and suggestions that helped us improve the paper. AN acknowledges the support of the Research Foundation - Flanders (FWO Vlaanderen). AG acknowledges support from INAF-Minigrant-2022 "LEGA-C" 1.05.12.04.01. FDE acknowledges funding through the ERC Advanced grant 695671 `QUENCH' and support by the Science and Technology Facilities Council (STFC). This research made use of Astropy,\footnote{\url{http://www.astropy.org}} a community-developed core Python package for Astronomy \citep{Astropy_2013A&A...558A..33A, Astropy_2018AJ....156..123A}. 
\end{acknowledgements}

\bibliographystyle{aa}
\bibliography{References}

\appendix
%

\onecolumn

\section{A comparison between the observed and modeled Lick indices} \label{ap:A}

\begin{figure*}[!ph]
    \centering
    \includegraphics[width=16.5cm]{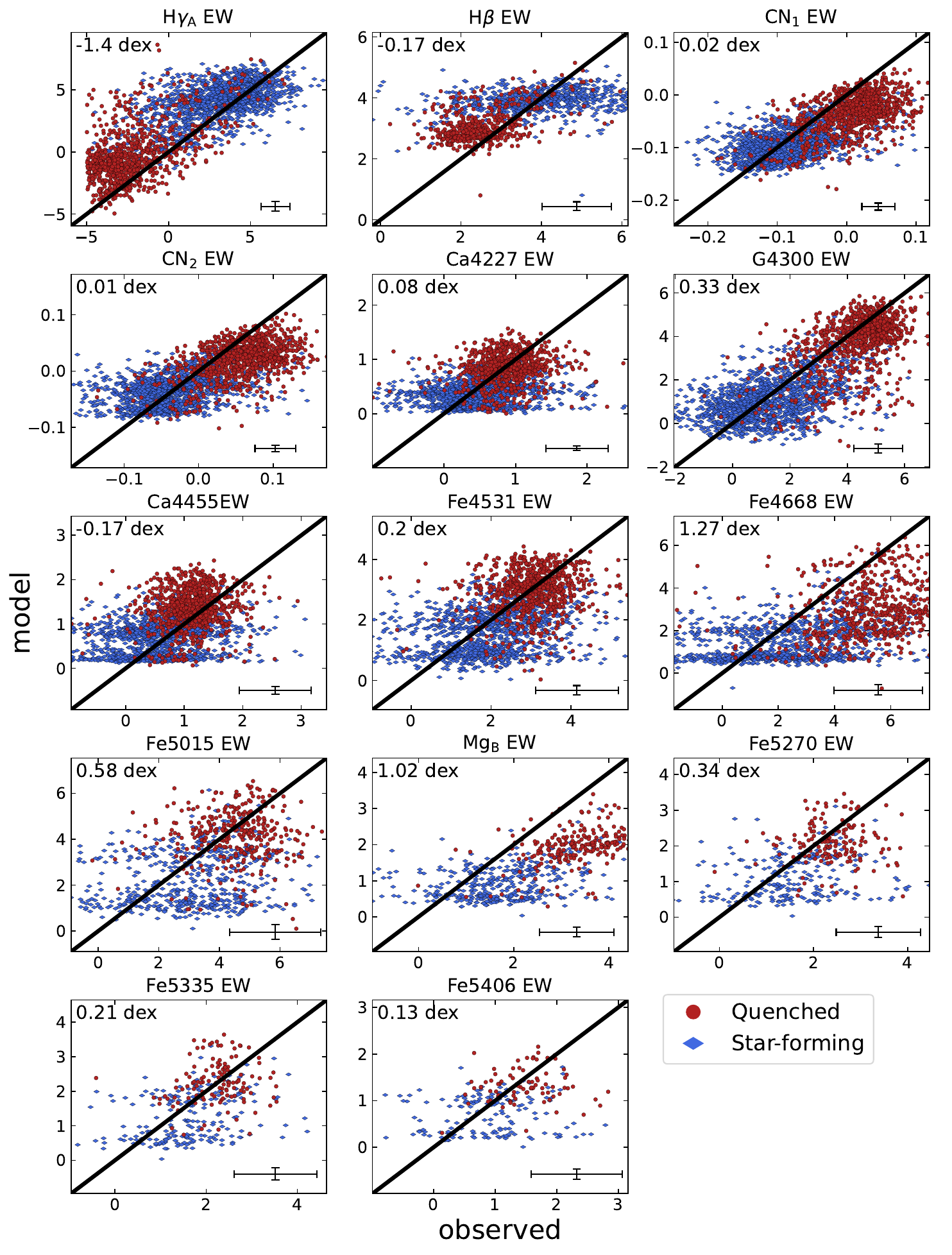}
    \caption{Observed absorption features compared to predictions from {\tt Prospector} fits to the COSMOS2020 photometry. Galaxies are color-coded by their $UVJ$--diagram classification to star-forming (blue diamonds) and quenched (red points). The absorption-only models are compared to the observed values (corrected for emission). The black line shows the one-to-one relation. The average offset on each Lick index is shown in the top-left corner of each panel.}
    \label{fig:a1}
\end{figure*}

Here, we present a comparison of 13 additional spectral absorption features, corrected for emission. The results of this comparison are shown in Fig.~\ref{fig:a1}. In each panel, the observed values of a spectral absorption feature are plotted on the \textit{x}-axis and the model values predicted from {\tt Prospector} fits to the COSMOS2020 photometry are plotted on the \textit{y}-axis. Galaxies are color-coded by their $UVJ$--diagram classification to star-forming and quenched. 

Similarly to the results shown in Fig.~\ref{fig:hd_fe4383_obs_vs_prd}, the global bimodality of star-forming and quiescent galaxies in photometric space is reproduced with the model spectra. However, we find that the majority of the model spectral features deviate considerably from their observed counterparts, with systematic offsets above 0.15~dex. This again implies that reliable age or metallicity determinations cannot be inferred from photometry alone.

\section{Mock analysis} \label{ap:B}

\begin{figure*}[t]
    \centering
    \includegraphics[width=18cm]{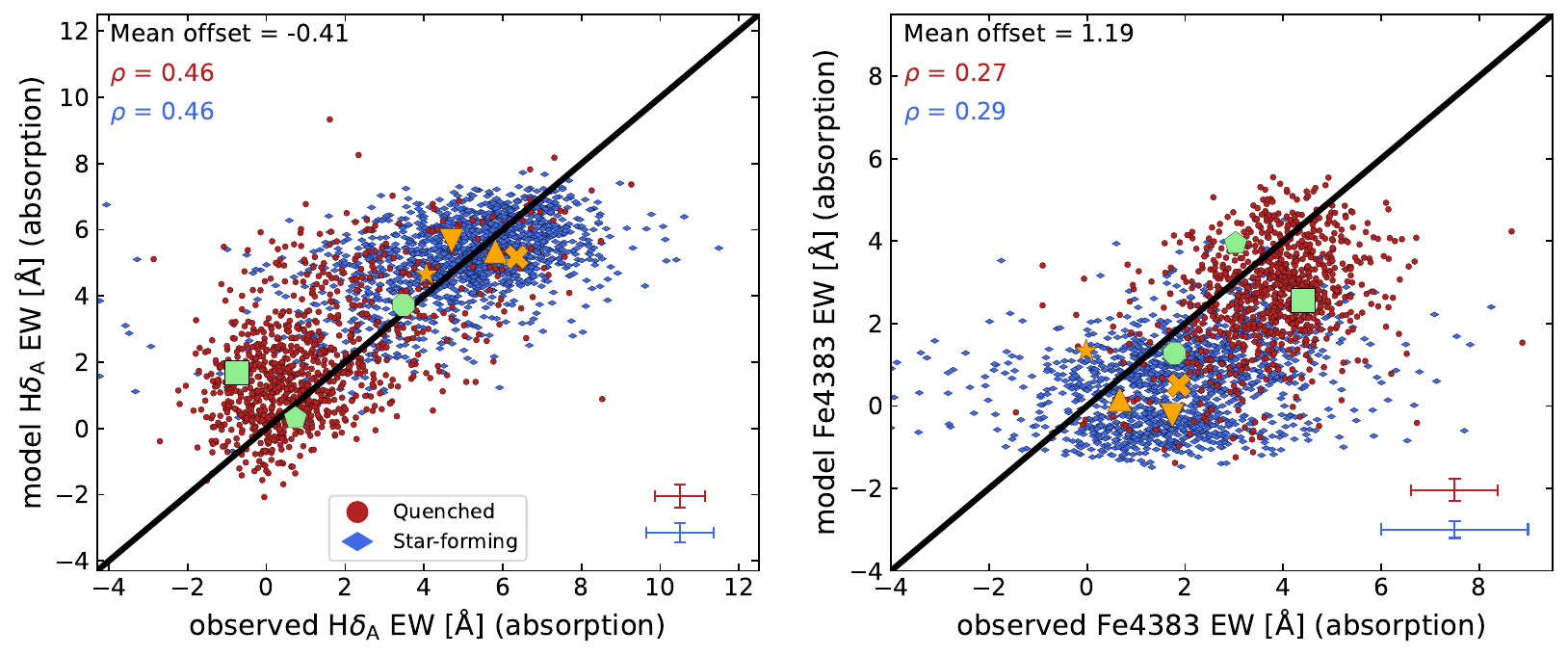}
    \caption{Observed \hda~and Fe4383 absorption features compared to predictions from {\tt Prospector} fits to the COSMOS2020 mock photometry. Galaxies are color-coded by their $UVJ$--diagram classification to star-forming (blue diamonds) and quenched (red points). The absorption-only models are compared to the observed values (corrected for emission). The black line shows the one-to-one relation. The various markers correspond to the star-forming (orange) and quenched (green) galaxies shown in Fig.~\ref{fig:sed_examples}. The statistics of the intrinsic scatter and the Spearman's rank correlation coefficient ($\rho$), are shown in the top-left corner of each panel. The median uncertainties of each axis and galaxy population are shown in the bottom-right corner of each panel.}
    \label{fig:mock_hd_fe4383_obs_vs_prd}
\end{figure*}

In this section, we wish to evaluate and explore possible biases in the resulted model spectra from our SED fitting method. First, we perturbed the original photometry of the COSMOS2020 catalog by introducing random noise, following a Gaussian distribution with $\sigma$ corresponding to the observed uncertainty per each photometric band. Then, we fitted the mock photometry with {\tt Prospector} and measured the Lick indices of the \hda~and the Fe4383 absorption lines from the mock spectra with {\tt pyphot}. The results are shown in Fig.~\ref{fig:mock_hd_fe4383_obs_vs_prd}. From this figure, we notice that the Lick indices from the spectra of the models that best fit the mock photometry are consistent with those we measured using the original COSMOS2020 photometry, with only slight differences in the average offsets. In other words, the main result of our analysis is not affected by any photometric biases.

\end{document}